\documentclass[12pt]{article}

\usepackage[letterpaper,hmargin=1in,vmargin=1in]{geometry}

\usepackage{graphicx,epstopdf,amsmath,amsfonts}

\def\be{\begin{equation}}
\def\ee{\end{equation}}
\def\ba{\begin{eqnarray}}
\def\ea{\end{eqnarray}}
\def\ge{\mathrel{\raise.3ex\hbox{$>$\kern-.75em\lower1ex\hbox{$\sim$}}}}
\def\la{\mathrel{\raise.3ex\hbox{$<$\kern-.75em\lower1ex\hbox{$\sim$}}}}

\def\simgt{\mathrel{\raise.3ex\hbox{$>$\kern-.75em\lower1ex\hbox{$\sim$}}}}
\def\simlt{\mathrel{\raise.3ex\hbox{$<$\kern-.75em\lower1ex\hbox{$\sim$}}}}

\newcommand{\bi}[1]{\bibitem{#1}}
\newcommand{\fr}[2]{\frac{#1}{#2}}

\newcommand{\nc}{\newcommand}

\nc{\gone}{\bar g_{\pi NN}^{(1)}}
\nc{\gzero}{\bar g_{\pi NN}^{(0)}}
\nc{\al}{\alpha}
\nc{\ga}{\gamma}
\nc{\de}{\delta}
\nc{\ep}{\epsilon}
\nc{\ze}{\zeta}
\nc{\et}{\eta}
\nc{\ka}{\kappa}
\nc{\rh}{\rho}
\nc{\si}{\sigma}
\nc{\ta}{\tau}
\nc{\up}{\upsilon}
\nc{\ph}{\phi}
\nc{\ch}{\chi}
\nc{\ps}{\psi}
\nc{\om}{\omega}
\nc{\Ga}{\Gamma}
\nc{\De}{\Delta}
\nc{\La}{\Lambda}
\nc{\Si}{\Sigma}
\nc{\Up}{\Upsilon}
\nc{\Ph}{\Phi}
\nc{\Ps}{\Psi}
\nc{\Om}{\Omega}
\nc{\ptl}{\partial}
\nc{\del}{\nabla}
\nc{\ov}{\overline}
\nc{\newcaption}[1]{\centerline{\parbox{15cm}{\caption{#1}}}}

\def\beq{\begin{equation}}
\def\eeq{\end{equation}}
\def\bmat{\begin{displaymath}}
\def\emat{\end{displaymath}}
\def\bear{\begin{eqnarray}}
\def\eear{\end{eqnarray}}
\def\ba{\begin{eqnarray}}
\def\ea{\end{eqnarray}}
\def\bery{\begin{array}}
\def\ery{\end{array}}
\def\bit{\begin{itemize}}
\def\eit{\end{itemize}}
\def\ben{\begin{enumerate}}
\def\een{\end{enumerate}}
\def\btab{\begin{tabular}}
\def\etab{\end{tabular}}
\def\btbl{\begin{table}}
\def\etbl{\end{table}}
\def\bfig{\begin{figure}[htb]}
\def\efig{\end{figure}}
\def\bpic{\begin{picture}}
\def\epic{\end{picture}}

\def\ga{\mathrel{\raise.3ex\hbox{$>$\kern-.75em\lower1ex\hbox{$\sim$}}}}
\def\la{\mathrel{\raise.3ex\hbox{$<$\kern-.75em\lower1ex\hbox{$\sim$}}}}
\def\gappeq{\mathrel{\rlap {\raise.5ex\hbox{$>$}}
{\lower.5ex\hbox{$\sim$}}}}
\def\lappeq{\mathrel{\rlap{\raise.5ex\hbox{$<$}}
{\lower.5ex\hbox{$\sim$}}}}

\def\gyr{{\rm \, G\kern-0.125em yr}}
\def\mev{{\rm \, Me\kern-0.125em V}}
\def\gev{{\rm \, Ge\kern-0.125em V}}
\def\tev{{\rm \, Te\kern-0.125em V}}

\begin{document}

\begin{titlepage}

\setcounter{page}{1}

\vspace*{0.2in}

\begin{center}

\hspace*{-0.6cm}\parbox{17.5cm}{\Large \bf \begin{center}

Resonant scattering and recombination\\ of pseudo-degenerate  WIMPs 
\end{center}}

\vspace*{0.5cm}
\normalsize

{\bf  Maxim Pospelov$^{\,(a,b)}$ and Adam Ritz$^{\,(a)}$}

\smallskip
\medskip

$^{\,(a)}${\it Department of Physics and Astronomy, University of Victoria, \\
     Victoria, BC, V8P 1A1 Canada}

$^{\,(b)}${\it Perimeter Institute for Theoretical Physics, Waterloo,
ON, N2J 2W9, Canada}

\smallskip
\end{center}
\vskip0.2in

\centerline{\large\bf Abstract}

We consider the direct and indirect detection signatures of WIMPs $\ch^0$ in 
kinematic regimes with a heavier, but nearly degenerate, 
charged state $\ch^{\pm}$.
For small splittings of ${\cal O}(10)$ MeV, the 
scattering of WIMPs off nuclei may be dominated by inelastic recombination processes mediated by the formation of $(\ch^- N)$ bound states, leading to a 
distinct signature for direct detection.  These cross-sections are bound primarily by limits on the abundance of heavy isotopes, 
and may be considerably larger than the elastic scattering cross section in more conventional models. If the mass splitting is  too large for recombination 
to occur, there may still be a significant  resonant enhancement of loop-induced electromagnetic form-factors of the WIMP, which can enhance the 
elastic scattering cross-section. We also discuss how this regime affects the annihilation cross-section and indirect detection signatures, and 
note the possibility of a significant mono-energetic $\gamma$-signal, mediated by resonant processes near the $(\ch^+\ch^-)$ bound state threshold.

\vfil
\leftline{March 2008}

\end{titlepage}

\section{Introduction}

The evidence for the existence of non-baryonic dark matter now comes from many sources and ranges over 
many distance scales \cite{review}, from the rotation curves of galaxies, the dynamics of clusters, lensing data, and the characteristics of
large-scale structure, to the features of the cosmic microwave background (CMB) fluctuation spectrum and the success of 
big bang nucleosynthesis (BBN). All of these pieces of astronomical data point to a similar cosmological density of 
dark matter, several times that of visible baryonic matter. Moreover, recent  observations of the bullet cluster \cite{bullet}
have for the first time pointed more directly to the existence of dark matter through spatial separation of the 
baryonic and non-baryonic (or dark) components in the collision.

This situation has been, and remains, one of the primary puzzles within particle physics and 
also one of the strongest motivations for physics beyond the Standard
Model. It is remarkable that the theoretical shortcomings of the Standard Model with regard 
to the UV sensitivity of the Higgs mass point to new physics at or near the
electroweak scale, and that a stable weakly interacting particle of this mass has just the right 
properties to be a dark matter candidate arising as a thermal relic from the Big Bang. There are also 
interesting links between various symmetries required to avoid large violations of approximate 
SM invariances, such as baryon number, and the stability required for such dark matter 
candidates. This apparent ``naturalness" of weak-scale cold dark matter, or WIMPs, has led to a vast 
literature on the subject in models such as the supersymmetric version of the Standard Model (MSSM) \cite{review}, models with
large extra dimensions \cite{hp}, etc, and also the development of ground-based direct-detection 
facilities searching for the recoil of large nuclei from elastic scattering with WIMPs in
the galactic halo \cite{cdms,xe}.

More recently, high precision data from WMAP \cite{wmap}, and the negative results of the LEP and Tevatron searches for new physics
at or near the electroweak scale has put some pressure on this seemingly natural scenario. 
In particular, in the MSSM with unification of the soft-breaking parameters at high energy scales (CMSSM), it is now well-known
that a generic point in the parameter space, consistent with collider constraints, 
would typically lead to a relic density for the neutralino LSP which is orders of magnitude too high, while 
only a rather tuned region of parameter space remains in agreement with data \cite{susy,review}. In fact, the simple freeze-out calculation of  the neutralino
WIMP relic density, which implies $\Om_\ch h^2 \sim 3\times 10^{-37}\,{\rm cm}^2/\langle \si_{ann} v \rangle$ \cite{LW},
leads to a value consistent with WMAP only in a region of the CMSSM parameter space which is already excluded by direct search bounds on 
the Higgs and chargino masses. Viable regions actually rely on additional, and to a certain extent accidental, features of the spectrum in order to 
enhance the neutralino annihilation cross-section. One such enhancement mechanism, the coannihilation \cite{GS} 
of the neutralino with a nearly degenerate charged slepton, allows the extension of the region 
of viable WIMP masses into the TeV energy domain.  This issue of tuning is of course not unique to the relic density of dark matter and is now a 
more generic problem for models of new weak scale physics following the absence of new discoveries at LEPII for example. 
Indeed, the relic density requirement and the apparent absence of new physics right at the weak scale 
suggests a prominent role for mechanisms which allow for enhanced annihilation of heavier WIMPs  
in coming years.

With this motivation in mind, in this paper we would like to focus on one particular enhancement 
mechanism, namely coannihilation with a 
nearly degenerate charged state, and explore how extreme limits of this 
kinematic regime may alter the predictions for direct and indirect searches. We will refer
to this kinematic scenario as a {\it pseudo-degenerate} WIMP, having 
in mind that the WIMP $\ch^0_1$ will be separated by a small mass gap of 
$\De m \sim {\cal O}(1-100)$ MeV from an excited state $\ch_2^0$ or $\ch_2^{\pm}$. 
For a WIMP mass of ${\cal O}(100\,{\rm GeV})$ this splitting is actually 
much smaller than is required for coannihilation, for which a relative splitting of 
less than 5\% is sufficient. Whether such near-degeneracies are
natural or not clearly depends on the precise nature of the model at hand, but it is a 
particularly interesting kinematic regime in the case that
$\ch_2$ is charged for the following reason. 
For $\De m < {\cal O}(20\,{\rm MeV})$, inelastic scattering of $\ch_1$ with heavy nuclei 
becomes possible in which capture of $\ch_2^{-}$ occurs to form
the bound state $(N\ch^-_2)$, and this process can dominate the cross-section if $\ch_1$ 
behaves in other respects as a conventional WIMP with
a suppressed elastic scattering cross-section. This process, since it involves 
charged exchanges, can lead to significantly different signatures in
direct detection experiments. If $\De m$ is larger than about 20 MeV, this recombination 
process is no longer possible, but there can still be a significant resonant
enhancement of the elastic scattering cross-section. Such enhancements of elastic 
scattering are in some respects the trade-off 
for going to a kinematic regime where the annihilation cross-section is similarly enhanced. 
For very small splittings, the annihilation cross-section
itself 
may also be dominated by resonant processes, since it is near threshold for the 
formation of the $(\ch^+\ch^-)$ bound state, which in turn
can result in an enhanced production of $\gamma$-rays. This is 
particularly so in the galactic 
environment where the characteristic velocities are quite low.

In what follows we will explore these issues in turn. In Section~2, we introduce the general kinematic scenario and some classes of
pseudo-degenerate models. In Section~3, we consider first the
recombination with nuclei and compute the simplest charged current and electromagnetic captures to the ground state, and comment on
various constraints from e.g. terrestrial heavy isotope searches. We then turn to the generic resonant
enhancement of the elastic cross-section. In Section~4, we consider the impact of resonant processes in annihilation,
and we conclude in Section~5 with some additional remarks on alternative detection signatures.

\section{Pseudo-degenerate WIMPs}

Following the motivation outlined in the preceding section, we will consider generic WIMP scenarios in which the dark sector has some
substructure, in the form of at least one excited state $\ch_2$ nearly-degenerate with the WIMP $\ch_1$ which we also take to have a mass
of ${\cal O}(0.1 - 1\,{\rm TeV})$, with the lower limit imposed by searches at LEPII. A near-degenerate neutral excited state was considered 
previously for different reasons in \cite{sm}, but for efficient coannihilation  
$\ch_2$ should have stronger interactions with the SM than $\ch_1$ and for the present paper we will generically assume that $\ch_2$ 
is electromagnetically charged, $\ch_2^\pm$, although a neutral state charged under SU(3)$_c$ would also fall into a similar class. These simple 
requirements define the class of pseudo-degenerate WIMPs to be studied in this paper.

\subsection{Models}

To provide a somewhat finer classification of pseudo-degenerate scenarios, we will generally consider two classes of models, determined
by the dominant interactions with gauge bosons. 
The WIMP $\ch_1$, whether fermion or scalar,  
is required to possess no diagonal tree-level spin-independent coupling to SM gauge bosons due to constraints on direct detection. Even a 
very small, $O(10^{-2})$,  
spin-independent coupling to the $Z$ is still sufficient to produce an elastic 
scattering cross-section with nuclei in excess of the
current bound. The nullification of couplings to $Z$ can be achieved either
 by a careful charge assignment in the WIMP sector, 
or by requiring that $\ch_1$ and its charge conjugate field $\ch_1^c$ are the same. 
In other words, in order to suppress the spin-independent scattering cross section, 
$\ch_1$ should be a real scalar if WIMPs have spin 0, 
or  a Majorana fermion if $\chi_1$ has spin 1/2.  
We have defined our scenarios such that $\ch_2$ is electromagnetically charged, and thus
the primary distinction we can place on the $\ch_1 - \ch_2$ WIMP sector is 
on the type of charged current between these two states. Restricting our discussion to tree-level 
couplings, we introduce two model classes.

\begin{itemize}
\item {\bf Type A --} The first scenario we will consider will allow for a charged current interaction with $W$-bosons. 
In other words, the pseudo-degenerate WIMP
sector possesses an off-diagonal vector current $J^\mu_{\ch_1\ch_2}$ so that,
\be
 {\cal L}_{\rm int} = J^{\mu -}_{\ch_1\ch_2} W^+_\mu + {\rm h.c.},
\ee 
and consequently $\ch_1$ and $\ch_2$ are either both bosons or fermions.
A familiar MSSM model in this class would involve a neutralino $\ch_1$ 
with a near-degenerate chargino $\ch_2^{\pm}$.
\item {\bf Type B --} The second scenario will be defined by the condition that $\ch_1$ and $\ch_2$ are respectively
a scalar and a fermion or vice versa. In this case the current between $\ch_1$ and $\ch_2$ is fermionic
and couples to the SM charged leptons $\psi$
\be
 {\cal L}_{\rm int} =  J^{a -}_{\ch_1\ch_2} \psi^+_a + {\rm h.c.}.
\ee 
A characteristic MSSM example in this case would be a neutralino $\ch_1$ with a 
near-degenerate  stau $\ch_2$.
\end{itemize}

In the remainder of this section, we outline the relevant kinematic regimes of 
interest.

\subsection{WIMP-nucleus binding energies}

Coannihilation does not impose undue levels of tuning on the mass spectrum; a splitting between the WIMP and excited states of ${\cal O}(5\%)$ is 
generally sufficient.  This does not significantly alter the expectations for interactions with baryonic matter, and we will be interested in a more
extreme limit in which the splitting is in the MeV range, and thus comparable to the Coulomb binding energy of $\ch_2^-$ with nuclei. We will defer any discussion of
theoretical motivations for this pseudo-degeneracy, but as a benchmark point to normalize the kinematic regimes
to be discussed below, it is worth noting that if $\ch_1^0$ and $\ch_2^{\pm}$ were truly 
degenerate in the dark sector through some symmetry, then the interaction  of $\chi_2^\pm$ 
with the SM gauge bosons would naturally imply a splitting of ${\cal O}(100\,{\rm MeV})$ 
\cite{strumia}.\footnote{It is also worth noting that if $\ch_1$ and $\ch_2$ arise from the low-energy hadronization of a single relic
state charged under SU(3)$_c$, e.g. as hadronized squarks \cite{eos}, they can naturally be split by ${\cal O}$(MeV) as members of an isospin doublet. In such
cases, recombination with nuclei could occur through strong rather than electromagnetic interactions, but this may be subject to more
stringent constraints than the scenarios for electromagnetic capture that we consider here. }

Denoting the mass difference as $\De m = m_{\ch_2} - m_{\ch_1}$, we explore the upper bound on $\De m$  
that leads to the formation of stable $(N\chi_2^-)$ bound states. In the limit $m_\ch \gg m_N$, 
the binding energy depends only on the nuclear mass,
and is naively given by the analogue of the Rydberg energy, $-E_b \sim (Z\al)^2 m_N/2$. 
More precisely, since the characteristic radius of the
orbit $r_B \sim (Z\al m_N)^{-1}$ generally lies well within the nucleus, we obtain a better estimate 
by solving the Schr\"odinger
equation with a given charge distribution inside the nucleus. This 
leads to the results shown in Table~1 for several elements that will be relevant in this 
paper. Two types of charge distribution with the same $\langle r_c^2\rangle $, Gaussian and step-like, are employed that can be viewed as two 
extreme approximations of a more realistic nuclear charge density.

\begin{table}
\begin{center}
\btab{||c|c|c|c||}
\hline
$(N\ch_2^-)$ & $Z$ &  -$E_b$ (MeV), Gaussian &  -$E_b$ (MeV), step-like \\
\hline
 $(^1$H$\ch_2^-)$ & 1 & 0.025 & -\\
 $(^4$He$\ch_2^-)$ & 2 & 0.35 & -\\
 $(  ^{11}$B$\ch_2^-)$ & 5 & 2.2 & 2.1\\
 $(^{12}$C$\ch_2^-)$ & 6 & 2.8 & 2.7\\
 $(^{14}$N$\ch_2^-)$ & 7 & 3.5 & 3.2\\
 $(^{16}$O$\ch_2^-)$ & 8 & 4.0 & 3.7\\
 $( ^{40}$Ar$\ch_2^-)$ & 18 & 9.1 & 8.0\\
 $( ^{74}$Ge$\ch_2^-)$ & 32 &  14.6 & 12.5\\
 $( ^{132}$Xe$\ch_2^-)$ & 54 & 21.7 & 18.4\\
\hline
\etab
\caption{\footnotesize Estimates for the 
binding energies of the state $(N\ch_2^-)$ assuming a gaussian and step-like nuclear charge distribution
for several relevant elements. }
\end{center}
\end{table}

Table~1 reveals three distinct kinematic regimes:
\begin{eqnarray}
  &\,&{\rm [(I):~recombination~with~light~elements]}\;\;\;\;\;\;\;\;\;\;\;\;\;\;0~ < \De m < 4~{\rm MeV}, \label{recom_range}
\\
  &\,&{\rm [(II):~recombination~with~heavy~elements]}\;\;\;\;\;4~{\rm MeV} < \De m \la 20~{\rm MeV}, \label{recom_range1}
  \\
  &\,&{\rm [(III):~resonant~loop~enhancement]}\;\;\;\;\;\;\;\;\;\;\;\;\;\;\;\;
20~{\rm MeV} \la \De m \la 100~{\rm MeV}. \label{res_range}
  \end{eqnarray}
In regime (I), stable bound states with light nuclei up to oxygen may be formed.  
Some models in this kinematic regime are significantly
constrained by searches for anomalously heavy isotopes of carbon and other light elements as 
 discussed below.
Regime (II) is equally interesting, as it opens the possibility of `recombination' processes where WIMPs
may combine with the heavy elements inside the detectors used for the direct searches of dark matter. 
Finally, kinematic regime (III) arises when $\De m$ is too large for recombination to occur in inelastic 
scattering, but still small relative to the WIMP mass, and the elastic scattering
cross-section can be resonantly enhanced through loop processes, 
with an off-shell $\ch_2^-$ appearing as an intermediate (virtual) state in scattering.
The upper limit in this case is not strict, but simply gives an estimate for the level beyond which these
processes are less relevant; 100 MeV is the characteristic centre-of-mass momentum for interactions of a 
100 GeV halo WIMP with terrestrial nuclei.

\section{Pseudo-degenerate WIMP-nucleus scattering}

\begin{figure}
\centerline{\includegraphics[bb=0 600 400 740, clip=true, width=10cm]{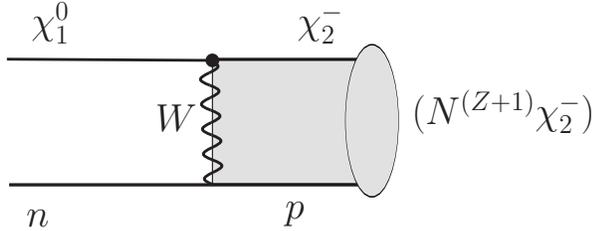}}
 \caption{\footnotesize The weak charged-current capture of $\ch_1^0$ to  form the
 bound state $(N\ch_2^-)$. }
\label{f1} 
\end{figure}

\subsection{WIMP-nucleus recombination}

One of the more interesting processes that becomes kinematically accessible 
for the mass splitting in (\ref{recom_range1}), is the `recombination'
of WIMPs with nuclei, and indeed such inelastic scattering can dominate 
the cross-section if $\ch_1$ is otherwise relatively inert, as would be
characteristic of a WIMP. In this section, we explore several classes of interactions.

\subsubsection{Weak charged-current recombination}

For models of type A, a
natural class of capture processes will proceed via weak currents, 
the simplest example of which is shown in Fig.~\ref{f1},
$\chi^0_1 + n \to \ch^-_2 + p$. Depending on the value of $\Delta  m$, the recombination process may occur 
directly to the ground state of the bound WIMP-nucleus system, or to an excited intermediate 
state that will subsequently decay  
to the ground state through emission of $\gamma$'s and/or neutrons depending on
the nucleus in question,
\be
 \chi^0_1 + N^{(Z)}  \to (N^{(Z+1)}\ch^-_2)^* \to (N^{(Z+1)}\ch_2^-) + (\gamma, n, \ldots). 
\label{wp1}
\ee
This process may receive a resonant enhancement but of course is also subject to the details of nuclear binding and so may not be 
energetically allowed for many light nuclei, depending on the relative binding energies of $N^{(Z)}$ and $N^{(Z+1)}$.
For example, the capture (\ref{wp1}) is not possible for such abundant nuclei as $^{12}$C, $^{14}$N, and $^{16}$O,
because the $(Z+1,A)$ nuclei are too massive. 
In addition to (\ref{wp1}), there always exists a $\beta^+$-type process, 
\be
\chi^0_1 + N^{(Z)}  \to (N^{(Z)}\ch_2^-) + e^++\nu,
\label{easy}
\ee
which is clearly non-resonant.

The non-resonant contribution (\ref{easy}) is relatively easy to estimate, 
and to refine to a full calculation 
if needed. For our purposes it suffices to estimate the non-resonant 
capture cross section by considering the overlap between the 
scattering and bound state wavefunctions, i.e.
\be
\sigma_{\rm non-res} \simeq \Gamma_{\chi_1\to \chi_2}
|\langle \ps_N^{\rm scat} | \ps_N^{\rm bs} \rangle |^2\times F_G,
\label{<>}
\ee
where $\Gamma_{\chi_1\to \chi_2}$ is the ``decay" width of $\ch_1$ to $\ch_2$
with the energy release $ Q_{\rm eff} = |E_b|-m_e-\Delta m$. Neglecting the Gamow factor $F_G$ for now,
saturating the wave-function overlap by characteristic nuclear scales, and taking 
$\Gamma_{\chi_1\to \chi_2}\sim  10^{-3}{\rm Hz}(Q_{\rm eff}/1{\rm MeV})^5$,  
we immediately discover that the rate for this process is extremely slow, 
\be
\sigma_{\rm non-res}v \sim 10^{-3}{\rm Hz} \times (2\,{\rm fm})^3 \times \left( \fr{Q_{\rm eff}}{1{\rm MeV}} \right)^5
\sim 10^{-51} {\rm cm}^2 \left( \fr{Q_{\rm eff}}{1{\rm MeV}} \right)^5,
\ee
even for the largest values of $Q_{\rm eff}$. 
 
Computing the cross-section for process (\ref{wp1}) with an intermediate excited nuclear state 
is rather nontrivial. However, to get a reasonable estimate, it is useful
to think of the excited state $(N\ch_2^-)^*$ as a resonance, for which Fig.~\ref{f1} 
characterizes the entrance width $\Ga_W$. Since $\Ga_W$ 
relates to a weak process, it is much smaller than the
decay width induced by electromagnetic and strong interactions, and we can use the Breit-Wigner formula
to write,
\be
 \si_{\rm res} \sim \sum_R \frac{\pi g_* }{q_{\rm cm}^2} \fr{\Ga_W\Ga}{(E-E_R)^2+\Ga^2/4}, \label{nwa}
\ee
where $g_*$ refers to the angular momentum multiplicity factor, 
$q_{\rm cm}$ is the momentum in the WIMP-nucleus centre-of-mass frame, and $E_R$ denotes the resonant 
levels of $(N\ch_2^-)$ which may be formed in the capture process.  

The spread in kinetic energy of the colliding WIMP-nucleus system is on the order of $M_R^2(N,\ch)v^2$,
where $v$ is the relative velocity and $M_R$ the reduced mass. Thus, for nuclear masses in the interval 
10 -- 100 GeV, the typical spread in kinetic energy is 10 -- 100 keV, which is somewhat smaller  
than the characteristic spacing of bound state energy levels of 1 MeV. This level spacing becomes 
even denser due to the additional Coulomb excitations in the $(N\ch_2^-)$ system, and for the purpose of 
obtaining an estimate it  is reasonable to assume that $\Delta E \sim 0.5\,$MeV. The decay width 
will vary quite significantly depending on $\Delta m$ and the nuclei participating in capture process. 
For electromagnetic decay widths in bound states with light nuclei it is reasonable to expect 
$\Gamma \sim O(1{\rm ~eV})$, while a neutron decay width in the capture of a WIMP with small 
$\Delta m$ by a large nucleus can easily reach $O(100~{\rm keV})$, and may thus be quite comparable 
to the spacing of energy levels.  For the latter case, we 
approximate the capture rate by assuming a characteristic energy denominator scale in (\ref{nwa})
on the order of $\Delta E$,
\be
 \si_{\rm res}v \sim 
\fr{1}{M_R^2v}\times \fr{\Ga_W}{{\rm  500 ~ keV}}. 
\label{simple1}
\ee

To complete the estimate we need to evaluate the weak entrance width. Its dependence on 
the main parameters in the problem is captured by the following scaling,
\be
\Gamma_W \sim  \frac{G_F^2 M_R^2v}{r_0^3},
\label{simple2}
\ee
where $r_0$ is the characteristic distance scale entering in the evaluation of the
matrix element, that for the purposes of this estimate we take to be of order the
nuclear radius of  4 fm, and we have also taken the coupling of $\ch_1\ch_2$ to $W$ to be {\cal O}(1).
The estimate (\ref{simple2}) may be viewed as the maximum possible 
weak rate, as it supposes a high degree of coherence among nucleons. 
Indeed, the $M_R^2$ dependence suggests a coherent response of all the nucleons
at characteristic momentum transfers of order $M_R v$, which might lead to an overestimate for large 
nuclei.  One finds that a parton-level calculation of the width, which effectively ignores 
coherence effects completely, leads to a result that is about an order of magnitude smaller.

Combining the two pieces together, (\ref{simple1}) and (\ref{simple2}), 
we arrive at an estimate of the capture rate given by 
\be
\si_{\rm res} v \sim \fr{G_F^2}{r_0^{3}\times ({\rm  500 ~ keV})}\sim 10^{-38}~{\rm cm^2}.
\ee
For light nuclei where the outgoing electromagnetic width is on the order of 1 eV, the rate is 
expected to suffer from a further suppression by $\Gamma_{\rm em}/\Delta E$, {\em i.e.} another 
five to six orders of magnitude. 
We observe
that the result is similar to the natural scale 
for this weak process, and thus does not appear to be significantly enhanced over more
conventional elastic scattering of $\ch_1$. However, we emphasize that this treatment of the nuclear
aspects of the calculation has been very cursory, and a more 
complete treatment may lead to further enhancement (or suppression) 
factors.

\begin{figure}
\centerline{\includegraphics[bb=0 600 400 740, clip=true, width=10cm]{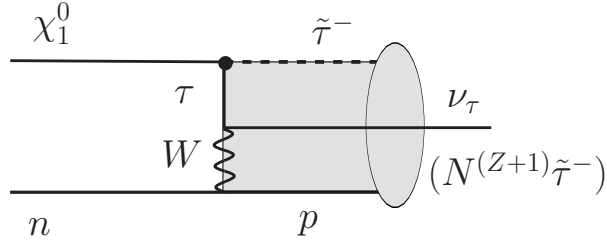}}
 \caption{\footnotesize Another charged current capture process of $\ch_1^0$ forming the
 bound state $(N\ch_2^-)$ and radiating a neutrino. }
\label{f2} 
\end{figure}

The capture process above is quite distinct from a characteristic nuclear recoil event due to elastic WIMP scattering because of the large 
${\cal O}$(MeV) scale energy release during the rapid decay of the excited bound state. Naively this makes it very difficult to observe 
in existing direct detection experiments which have detectors tuned to detect a fiducial recoil energy of no more than 100 keV. For this
reason, it is interesting to consider inelastic processes that would have a signature somewhat closer to a standard recoil event, and
one possibility is shown in Fig.~\ref{f2}, for the special case in which $\ch_2$ is a scalar partner of the $\ta$ as in the MSSM. 
The reverse process, a nuclear-assisted stau to neutralino decay was discussed previously in \cite{Jittoh}. 
Such a capture process
is analogous to (\ref{wp1}), with the important distinction that the bound state de-excites through radiating neutrinos which would escape the 
detector leaving just the recoil signal to be observed. However, this process is further suppressed and may not significantly impact 
the cross-section purely through elastic scattering of $\ch_1$.

\subsubsection{Electromagnetic recombination}

\begin{figure}
\centerline{\includegraphics[bb=0 500 600 740, clip=true, width=12cm]{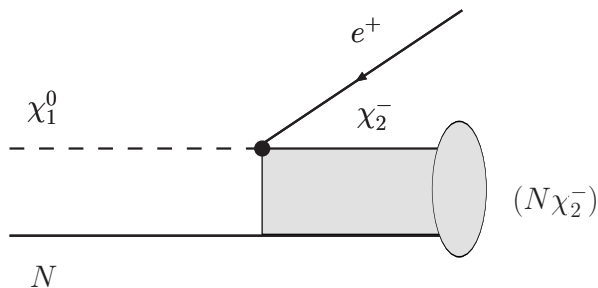}}
 \caption{\footnotesize The electromagnetic capture of $\ch_1^0$ to form the
 bound state $(N\ch_2^-)$ and radiating a positron. }
\label{f3} 
\end{figure}

For models of type B, which lack a weak charged current, alternative recombination channels are open
 if the charged states represent bosons and the neutral states fermions, or vice versa. In particular,  the 
 electromagnetic processes shown in Fig.~\ref{f3} become possible. In the early universe 
the abundance of charged states is then rapidly depleted via processes such as $\chi^\pm_2 \to \chi^0_1+ e^\pm$, 
while in the current epoch $\ch_2^-$ may be regenerated via recombination in the form,
\be
\chi^0_1+ N \to (N \chi^-_2) +e^+,
\label{X0capture}
\ee
which may occur given the appropriate kinematics (\ref{recom_range},\ref{recom_range1}),  where $N$ is again 
a generic nucleus, and $(N \chi^-_2)$ a stable bound state. Process (\ref{X0capture}) may appear to violate lepton number, and 
thus be suppressed. However, this need not be the case if either $\ch_1$ or $\chi_2$ carry lepton number as happens e.g.
in models where $\chi_1$ is a sneutrino, and $\chi_2$ a chargino, or when $\chi_1$ is a neutralino, and
$\chi_2$ a charged slepton. In the widely discussed neutralino-stau scenario, the coupling to the positron would actually 
represent a flavour-changing process in the lepton sector, which may be somewhat suppressed but does not have to be 
vanishingly small. 

Process (\ref{X0capture}) is an interesting variant of standard radiative recombination of a nucleus with an electron, 
in that Coulomb interactions are
present in the final rather than the initial state, and since the positron has to tunnel 
out of the nucleus the cross-section will be Gamow-suppressed,
particularly when the positron is non-relativistic. 
To be concrete, we will consider the situation where $\ch_1$ is a scalar and $\ch_2$ a Dirac fermion, with the effective vertex,
\be
 {\cal L} = g \ch_1 \bar{e} \ch_2^{-} + {\rm h.c.}
\ee
The free decay width $\Ga_{\ch_2}$ of $\ch_2$ in the early universe is then sufficiently fast to avoid problems with 
BBN provided $g^2 > 10^{-17}$ \cite{CBBN1}.

We estimate the capture rate by considering the overlap between the scattering and bound state wavefunctions 
as in (\ref{<>}), i.e.
$\sigma \simeq \Gamma_{\chi_1\to \chi_2}
|\langle \ps_N^{\rm scat} | \ps_N^{\rm bs} \rangle |^2\times F_G$ 
where $\Ga_{\chi_1\to \ch_2}$ is the free decay width (for $\De m <0$), and $F_G$ is the Gamow factor associated with the interaction of the radiated positron
with the nuclear potential. In the centre-of-mass frame of 
$\ch_1$ and $N$, the positron energy is $E_{e^+} \approx |E_b| + |E_{\rm kin}| - \De m + m_e \sim$ 5-15 MeV, 
while the nuclear barrier is characteristically
of ${\cal O}(Z\al/R_N)$ that may reach 20 MeV for large nuclei. In that case the positron will have to 
tunnel out of the Coulomb barrier leading to a suppression in 
the cross-section, i.e. $F_G \ll 1$. We will discuss this point further below.

In what follows we consider the capture to the ground state that maximizes the kinetic energy of the outgoing 
positron and thus minimizes the Gamow factors.  
The bound state wavefunction can be taken as that of a 3D harmonic oscillator,
with the inner non-Coulombic 
part of the nuclear potential given by $V(r) = (Z\al)/(2R_N)\left( (r/R_N)^2 - 3\right)$. 
Computing the overlap and converting to nuclear parameters, we obtain
\be
 \langle \si_{\rm rec} v \rangle = \frac{g^2}{2\pi} \frac{(E_{e^+} + m_e)|p_{e^+}|}{M_{\ch}} 
R_N^3 \left(\frac{a_B}{R_N}\right)^{3/4} 8\pi^{3/2} \exp\left(-(\mu v)^2 R_N^{3/2} a_B^{1/2}\right) \times F_G, \label{cap}
\ee
where $\mu=M_R(N,\ch)$ is the reduced mass of $N$ and $\ch_1$, $a_B=1/(Z\al \mu)$ is the `Bohr radius', and $R_N$ the nuclear radius.

If we suppress the detailed dependence on the bound state wavefunction, then  in the limit of a relativistic positron 
and assuming $a_B\sim R_N$, this reduces to
  \be
  \langle \sigma_{\rm rec} v \rangle \sim g^2 a_B^3 \fr{E^2_{e^+}}{ M_\chi},
  \label{modelrate}
  \ee
which exhibits the inverse scaling of the inelastic cross-section with the WIMP momentum. The most significant numerical 
correction to (\ref{modelrate}) arises from the fact that in practice $a_B \ll R_N$ for the nuclei of interest. The exponential factor 
in (\ref{cap}), associated with the ground state wavefunction, provides a correction of ${\cal O}$(20\%) for large nuclei.
Evaluating the estimate (\ref{modelrate}) for WIMP capture on light nuclei, such as carbon or nitrogen, 
we take $a_B \sim 1$ fm, and $E_{e^+} \sim $ 1~MeV, 
to obtain the recombination rate
\be
\langle \sigma_{\rm rec} v \rangle_{\rm light\, nuclei} \sim 10^{-34} ~{\rm cm^2} \times \left(\fr{g^2}{4\pi\alpha_w}\right) \times \left(\fr{1~{\rm TeV}}{M_\chi}\right).
\ee

For heavier nuclei, the Gamow factor $F_G$ can no longer be ignored and leads to significant suppression if the 
positron energy is well below the nuclear potential barrier. The present situation differs slightly from conventional non-relativistic tunneling,
where $F_G \sim e^{-G}$ with $G = \int_C dr |k|  \sim 2\pi Z\al/v$,
since for generic values of $\De m$ the positron will be relativistic even for energies well below the barrier. Consequently, $Z\al/v$ will not be too large
and we can ignore Coulomb resummation effects after tunneling has occurred and simply account for the tunneling factor.
This we estimate as  $G^{\rm rel} = {\rm Re} \int_C dr |k_b|$ using the relativistic momentum $k_b$ under the (wide) nuclear potential, which leads to
$G^{\rm rel} \sim 2\pi Z\al E/k$ provided the positron is only mildly relativistic. The latter constraint is necessary to be in the ``wide barrier'' regime which 
restricts the positron to have $\gamma \la {\cal O}(5)$.
For the values of $Z$ relevant here, the cross-section is still somewhat suppressed for generic $\gamma$-factors of the emitted positron, but far less
so than in the non-relativistic regime.
The impact is most significant for our maximal choice of $\De m$ as 
the positron is then barely relativistic even with capture to the ground state, and Coulomb effects may also become important. For 
a radiated positron with $\gamma\sim 1-2$, we find $F_G \sim 10^{-3}-10^{-1}$.

Inserting characteristic numerical scales for heavy nuclei into (\ref{cap}), taking $E_{e^+}\sim 10$~MeV and setting $F_G\sim 10^{-2}$, we can convert this into 
an effective cross-section per nucleon, the characteristic quantity quoted in direct WIMP searches,
\be
 \si_{p,n} = \frac{M_R^2(p,\ch)}{M_R^2(N,\ch)} \frac{1}{A^2} \frac{\langle \si_{\rm rec} v\rangle}{\langle v \rangle} \sim 10^{-39} \left(\frac{g^2}{4\pi \al_w}\right)\, {\rm cm}^2,
\ee
where $M_R$ denotes the reduced mass, and which for $g^2/(4\pi) \sim \al_w$ is 
interestingly a few orders of magnitude above the current direct 
detection constraint on elastic scattering, e.g. with Ge or Xe \cite{cdms,xe}.

Its also worth noting that an additional inelastic channel is triple recombination with the use of a $K$-shell electron,
\be
\chi^0_1+ e^- + N \to (N \chi^-_2)^* \to (N \chi^-_2) + \gamma.
\ee
This process does not require tunneling through the potential barrier, but is suppressed by the 
small probability of finding the $K$-shell electron near the nucleus, $(Z\alpha m_e R_N)^3$. This channel is expected to remain 
subdominant except for the case of very small $E_{e^+}$, {\em i.e.} with $\Delta m$ very close to the capture threshold.

\subsubsection{Constraints from BBN and heavy isotope searches}

As is well-known in the context of charged dark matter, there are stringent constraints on the presence of long-lived charged particles
surviving from the Big Bang. It has recently been recognized that charged states with 
lifetimes longer than ${\cal O}(10^3\,{\rm s})$ can also severely
disrupt the successful predictions of BBN through the catalysis of interactions that, for example, 
would overproduce $^6$Li and $^9$Be by many orders of
magnitude \cite{CBBN1,CBBN2}. For this reason, we will simply assume here that the state $\ch_2^{\pm}$ 
has a sufficiently small lifetime to decay before 
nucleosynthesis:
\be
 \ta_{\ch_2} < {\cal O}(1) \,{\rm s}. 
 \label{lifetime}
\ee
This can be a significant constraint on realizations of this scenario within the MSSM, 
where for example $\ch_2 = \tilde{\ta}_1$. In the absence of a flavour-changing 
 stau-neutralino-electron coupling, the condition (\ref{lifetime}) is equivalent 
 to requiring that the mass splitting between stau and neutralino is in excess of 100 MeV. 
 However, even an extremely small coupling to the electron, $g \sim O(10^{-8})$ would allow 
 (\ref{lifetime}) to be satisfied without imposing significant constraints on $\Delta m$. 

More direct constraints on the mass splitting $\De m$ then arise from 
the terrestrial limits on the abundance of anomalous heavy isotopes. Here the strongest
constraints apply to light elements, which are in some sense also the most reliable, arising from well-mixed gaseous
or liquid media. The most stringent is the abundance constraint on heavy hydrogen (He$\ch_2^-)$, in the form of 
heavy water \cite{smith82}, which however is
only relevant here if the mass splitting is very small, less than 350 keV. There 
are also quite stringent constraints on (C$\ch_2^-)$ 
(relative abundance $f_B < 10^{-14}$ per nucleon) and (N$\ch_2^-)$ ($f_C < 10^{-20}$ per nucleon)
\cite{hemmick90}.

To make use of these constraints, we connect the abundance of the parent nuclei that undergoes the capture process 
to that of the anomalously heavy daughter nucleus, a bound state with $\chi_2$,
\be
\fr{ n_{\rm daughter} }{n_{\rm parent}  } = \langle \si_{\rm rec} v \rangle n_{\rm DM} \ta_{\rm exp}.
\ee 
Here $\ta_{\rm exp}$ is the exposure time, and $n_{\rm DM}$ is some 
 average dark matter density near the solar system, which we shall assume to be equal to the local WIMP
number density, $ n_{\rm DM} \sim 3\times ({\rm TeV}/m_{\chi})\times 10^{-4}~ {\rm cm}^{-3}$. 
Specializing to the case of  electromagnetic capture (\ref{X0capture}) on nitrogen, 
and using the limit on the anomalous carbon abundance, $f_C < 10^{-20}$ per nucleon, we arrive 
at the following parameter constraint, 
\be
 \left( \fr {1~\rm TeV}{m_{\chi}} \right)^2 \times \left(\fr{g^2}{4\pi\alpha_w^2}\right) < 10^{-9}\;\;\;\;\;\; \mbox{for Type B models with}\, \De m \la 3\,{\rm MeV}.
\label{anomalousC}
\ee
In deriving this constraint we also assumed that carbon and nitrogen are almost equally abundant in nature,
which is rather conservative: the carbon used in Ref.~\cite{hemmick90} presumably spent the majority of its time in
the atmosphere, where nitrogen is significantly more abundant. 
Nonetheless, the constraint (\ref{anomalousC}) is very significant, and for example implies the absence of a 
large coupling to the first generation of leptons in the neutralino-stau model. 
It clearly disfavors models of type B with mass splittings below $\sim 3$~MeV, when the binding to nitrogen 
becomes energetically possible.

It is remarkable that the abundance constraints on anomalous
isotopes of heavier elements are considerably weaker, while there is also
greater uncertainty over whether the samples tested have a characteristic exposure time. 
Even interpreted in the most conservative manner,
existing bounds e.g. on heavy isotopes of Au \cite{mohapatra}, from (Hg$\ch_2^-$), lead to 
relatively mild constraints that are well below the
estimates for the cross-sections obtained in this section. For this reason, the 
range quoted in (\ref{recom_range1}) seems perfectly
viable, although it would of course be interesting were further searches to be 
performed for heavy isotopes of elements with $Z>10$.

Thus,  we conclude that for models of Type B heavy isotope searches do impose a significant constraint but only in connection
to binding with light elements and do not overly constrain the cross sections for $\De m$ in the range (\ref{recom_range1}). 
For models of Type A, where only weak charged currents are allowed, we noted earlier that the nuclear binding energies do not
permit capture by the light C, N, and O nuclei. Thus, since the 
capture rates are too slow to provide a significant constraint from 
searches for anomalous isotopes of heavy nuclei, the entire range of $\Delta m$ including the 
smallest values down to an MeV is allowed. We note in passing that models with weak currents, 
$\Delta m\,\sim\;$few MeV, and 
the $\chi_2^-$ lifetimes of $\sim 2000$ seconds alleviate the known problems with the primordial abundance of 
lithium \cite{CBBN2}.

\subsubsection{Constraints from direct detection and from annihilation in the sun}

The existing limits on the elastic cross-section per nucleon have reached impressive 
levels of order $10^{-43}$~cm$^2$ for a 100~GeV WIMP and
further progress is anticipated in coming years. However, its worth bearing in mind that the detection strategy in this case relies crucially
on a characteristic recoil signal with energy of order 50 keV. 
In this context, its interesting to consider if direct terrestrial dark matter searches are indeed sensitive
to electromagnetic recombination processes of the type shown in (\ref{X0capture}). 
At first sight, it is unclear if the recoil during the capture process would be observed, 
due to the additional MeV-scale energy release associated with 
the subsequent positron annihilation in the detector. For this reason, we cannot
immediately place constraints on the size of the coupling 
$g$ from this source. Indeed, for the majority of detectors, events with such a large energy 
release would be identified with background and rejected. 

Besides the direct detection strategy that is currently focused on the search for WIMPs elastically scattering off nuclei, 
another traditional method to look for WIMPs is by searching for highly energetic neutrinos coming from  WIMP annihilation in the
solar core. The density of WIMPs in the solar core can be many orders of magnitude larger
than $n_{\rm DM}$ due to gravitational capture through down-scattering of WIMPs on nuclei inside the sun, with 
an efficiency again controlled by the elastic scattering cross section. However, the pseudo-degenerate WIMP regime can significantly modify 
the sensitivity of this method. Indeed, if $\Delta m$ is less than about $10$ MeV, sufficiently small to allow recombination with
iron, a trapped WIMP will preferentially undergo iron capture rather than participate in direct annihilation 
with another WIMP. This stems from the fact that the density of iron in the solar core 
is many orders of magnitude larger than the density of WIMPs, and that 
the capture rate, (\ref{simple2}) or (\ref{modelrate}), is quite comparable to or even larger than the annihilation 
rate. Simple estimates show that for a typical capture rate 
of $\langle\sigma v\rangle \sim 10^{-40}{\rm cm}^2$,  a trapped WIMP recombines with an iron nucleus within $10^{8}$ seconds which 
is a much shorter time scale than for annihilation with another WIMP. Therefore, most of the WIMPs inside
the sun would be in bound states with iron and therefore shielded from annihilation by a large Coulomb barrier, 
which would essentially cut off the flux of ultra-energetic neutrinos.

\subsection{Resonant enhancement of EM form-factors}

The characteristic momentum for an ${\cal O}$(100 GeV) WIMP in the galactic 
halo scattering with nuclei is $m_\ch v \sim$ 100 MeV, which implies that even if the
mass splitting $\De m$ between $\ch_1$ and $\ch_2$ is too 
large to allow for recombination with nuclei, significant resonant enhancements
from the off-shell $\chi_2$ 
are possible in the elastic scattering cross-section. This is the 
scenario we will explore in this section, having in mind a splitting in the range
(\ref{res_range}), so that $\ch_2$ appears off-shell with virtuality $q^2 \sim m\Delta m$ 
in the scattering process as shown schematically in Fig.~4. 

\begin{figure}
\centerline{\includegraphics[bb=0 520 600 700, clip=true, width=12cm]{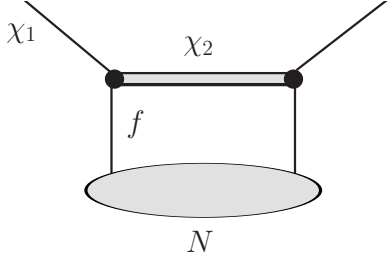}}
 \caption{\footnotesize A schematic representation of the resonant effects we will consider, where the shaded region denotes the interaction with the nucleus. }
\label{f4} 
\end{figure}

The precise nature of the amplitude will depend on the mechanism 
via which the WIMP interacts with the nucleons. The general $\ch_1 N \rightarrow \ch_1 N$
elastic scattering amplitude will involve many contributions, but since the WIMPs in the galactic halo are non-relativistic, one can summarize the dominant
contributions through the leading low-$q^2$ components of form-factors of the WIMP with various currents to which the nucleon couples.
Imposing minimal assumptions on $\ch_1$, namely the absence of tree-level vector couplings to $\gamma$, $Z$, and gluons, there are a number of 
form-factors which may play a role in interactions with the nuclei in the detector. To keep the discussion manageable, we will restrict our attention
to electromagnetic form-factors, although one should bear in mind that much of the discussion will translate directly to couplings to other SM
gauge bosons, and in various scenarios gluonic or weak form-factors may be of more importance. It is also important to keep in mind that these
form-factors are often only well-defined off-shell and thus in practice this is just a convenient means of keeping track of the leading parts of the 
scattering amplitude in an appropriate kinematic regime. 

Recall that for a {\it neutral} state $\ch$ of arbitrary (nonzero) spin, if we impose $T$-invariance, we can expand the 
matrix element of the electromagnetic current  \cite{kp} as follows in powers of the momentum transfer $q=p-p'$, \footnote{For particles with no vector coupling to the
photon, the anapole moment $a$ (and also the charge radius $r_D^2$) are strictly defined in a gauge invariant manner only through an external current.
They determine a coupling to an off-shell photon, as is relevant to nucleon scattering, and one should also include the 
corresponding off-shell $Z$ coupling for gauge invariance. We will ignore this issue as in practice for the momentum scales relevant here
this effect is negligible.}
\ba
\langle p'|J_\mu | p\rangle_{T\; {\rm even}} &=& e\bar{\ch}_1(p') \left[ -\frac{1}{6} r_D^2 q^2\gamma_\mu  
+  a(q^2\et_{\mu\nu}-q_\mu q_\nu)\gamma_5 \gamma^\nu -i\mu \Si_{\mu\nu}q^\nu \right. \nonumber\\
  && \;\;\;\;\;\;\;\;\;\;\;\;\;\;\;\; \left.- \frac{1}{4}Q(Sq)^2\gamma_\mu + \cdots \right] \ch_1(p),
\ea
where $\Si_{\mu\nu}$ is the spinorial generator in the appropriate representation (i.e. $\si_{\mu\nu}$ for a spin-1/2 particle) and
the corresponding spin 4-vector is $S^\mu = i \ep^{\mu\nu\rh\si} \Si_{\rh\si} q_\nu/(4m_\ch)$. The small-$q^2$ limit allows an expansion of the general form-factors
and the terms on the first line contain the leading
moments for a spin-1/2 particle -- the charge radius $r_D^2$, the anapole moment $a$ which is $P$-odd, and the magnetic dipole moment $\mu$. For spin $S\geq 1$,
there are also higher order multipoles and the second line contains the leading term in this limit, namely the electric quadrupole moment $Q$. 
Many of these moments vanish given specific constraints on $\ch_1$, while on the other hand if we
relax the constraint on $T$-invariance, there are in addition the electric dipole moment, the magnetic quadrupole moment, etc. but these will not be
needed in what follows.
  However, particularly in the case of scalar particles where (with the exception of an appropriately defined charge radius) these moments do not exist, we 
 need to go beyond one-photon exchange and consider two-photon processes. We will again consider the small $q^2$ limit, in which case the leading constant 
 parts of the form-factors are the electric and magnetic susceptibilities $\et_E$ and $\et_B$, and write the
 matrix element in position space,
 \be
   {\cal L}_{2\gamma} = \bar{\ch} F^{\al \mu} F^\beta_\mu \left[ \frac{1}{2m_\ch^2}(\et_E+\et_B) \ptl_{\al}\ptl_{\beta} + \frac{1}{4}\et_B\et_{\al\beta}\right]  \ch + \cdots
    \ee
 Once again, if we allow for $T$-violation, there are also mixed polarizabilities given by $\bar{\ch} F\tilde{F} \ch$, but we will drop these
contributions in what follows.

At the non-relativistic level, the relevant constant coefficients in the form-factors are conveniently assembled into a Hamiltonian
describing the interaction of the non-relativistic WIMP with a slowly varying electromagnetic field. The general form of this Hamiltonian
was considered previously in \cite{ptv}, and with the restriction to $T$-invariance as above, takes the form,
\ba
 {\cal H}_{\rm nr}^{T\; {\rm even}} &=& - \mu {\rm \bf B}\cdot \hat{\bf S} - a {\bf j} \cdot \hat{\bf S}  -  \frac{1}{4}Q_{ij}\ptl_i E_j \nonumber\\
   && \;\;\;\;\;  - \frac{1}{6} e r_D^2 \del\cdot {\bf E} - \frac{1}{2} \et_E E^2 - \frac{1}{2} \et_B B^2 + \cdots
    \ea
  with $Q_{ij}= Q(S_{(i} S_{j)} - (2/3)\de_{ij} S(S+1))/(S(2S-1))$, in terms of the 
various multipole moments and susceptibilities above.
  The moments on the first line apply only for nonzero spin. If the particle is a 
Majorana fermion then furthermore $Q_{ij}=\mu=r_D^2=0$. The vanishing of the effective charge 
radius of  a WIMP has significant consequences for WIMP-nucleus scattering, as $r_D^2$ 
contributes to spin-independent scattering enhanced by a very large nucleon coherence factor.

\subsubsection{Resonant enhancement of loop processes}

\begin{figure}
\centerline{\includegraphics[bb=0 500 600 740, clip=true, width=8cm]{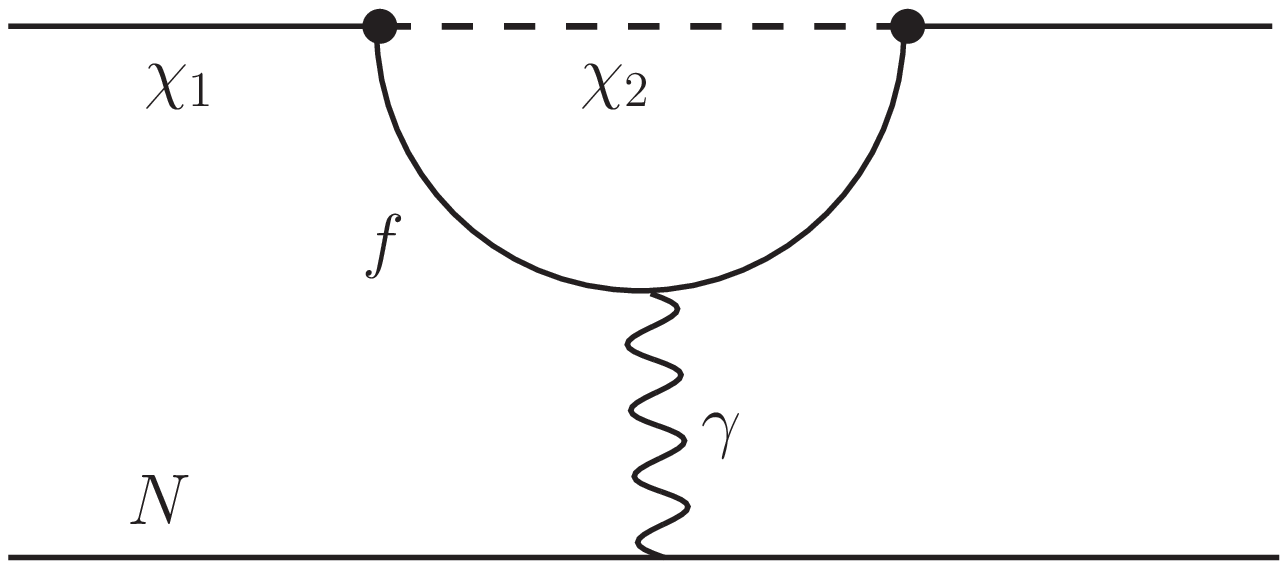}\includegraphics[bb=0 500 600 740, clip=true, width=8cm]{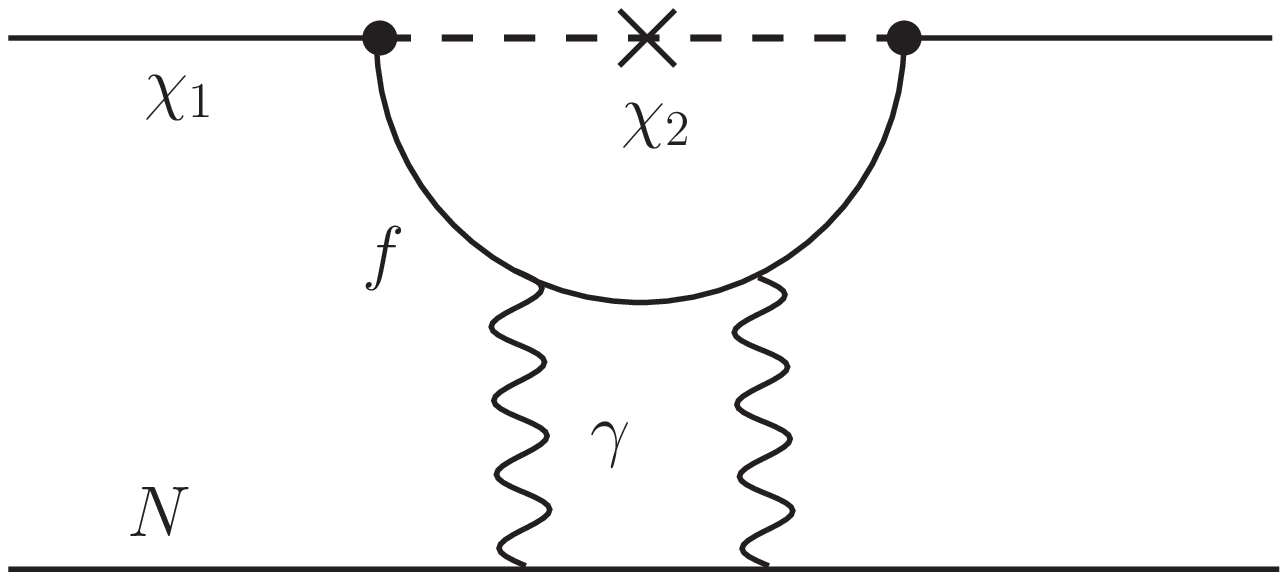}}
 \caption{\footnotesize A contribution to the scattering cross-section of $\ch_1$ with a nucleon, mediated by the  multipole moments (on the left) or susceptibilities (on the right). }
\label{f5} 
\end{figure}

The cross-sections for non-relativistic scattering due to these moments are listed for example in \cite{ptv}. The main point we wish to emphasize here is that even if these
moments are zero at tree-level, for pseudo-degenerate WIMPs they naturally arise at loop-level and indeed may be resonantly enhanced. The characteristic
diagrams are shown in Fig.~\ref{f5}.

Let us focus first on models of type B, where the WIMP $\ch_1$ is a fermion with a nearly degenerate scalar partner $\ch_2$ which
appears resonantly in the scattering process. 
Its convenient to denote the ``moments'' generically as $M_{(d)}$ where $d$ is the mass dimension of the
corresponding operator, i.e. $d=1$ for $\mu$, $d=2$ for $r^2_D$ and $a$, while $d=3$ for $\et_E$. An inspection of the loop diagrams indicates
that the decoupling with the WIMP mass $m_\ch$ is generic, i.e. $M_{(d)} \sim 1/m_\ch$, given that the extra internal fermion state is light, consistent
with resonance. This scaling can be straightforwardly understood by taking the non-relativistic limit for the heavy fermions. This leads to a factor 
of $\sqrt{m_\ch}$ at each vertex, while the heavy scalar propagator leads to a factor of $1/m_\ch^2$. Moreover, in the resonant limit the
loop is infra-red divergent and thus the result is enhanced by the small mass scale $\De m$ which sets the cutoff momentum scale, provided a light
SM fermion with $m < \De m$ is available. Otherwise, one must replace $\De m$ with the corresponding mass. E.g. in the case of a near neutralino-stau degeneracy,
the infra-red cutoff is fixed by $m_\ta$. Thus in the generic case, the moments scale in the following way in the resonant limit,
\be
 M_{(d)} \sim \frac{({\rm loop\;factor})}{m_\ch} \times \frac{1}{(\De m)^{d-1}}, \label{scaling}
\ee
where in fact the loop factor is also enhanced by $\pi$ due to the infrared divergence. 
This formula is quoted for zero momentum transfer, and in practice 
 for  $q^2 \ga m_\ch\Delta m > (m_\ch v)^2$ the EM form-factors will 
decrease as a function of  $q^2$. 
This scaling suggests that the enhancement is all the
more impressive for the susceptibilities, but one must bear in mind that these 
amplitudes are ${\cal O}(\al^2)$ and more importantly  involve a chirality flip and 
so require a double degeneracy of $\ch_1$ with both $\ch_{2L}$ and $\ch_{2R}$.

We have studied the contributions of these enhanced amplitudes to the scattering cross-section and, although significant, the results in most cases
do not reach the level of the current direct search bounds. For this reason, we will not present full formulae, but simply comment on a few interesting
cases. 

\begin{itemize}
\item {\bf Charge radius --} The exception to this general conclusion involves the charge radius in any scenarios in which it is present at 1-loop, arising e.g. from
the diagram on the left of Fig.~\ref{f5}. A straightforward calculation leads to a result of the form $r_D^2 \sim \al/(m_\ch \De m)$ where the order-one normalization
depends on the precise field content. For $\De m \sim {\cal O}(100~{\rm MeV})$ this leads to a (per nucleon) 
elastic scattering cross-section \cite{ptv} on $^{72}$Ge for example of
order $10^{-38}$ cm$^2$ which is several orders of magnitude above the current bound, and close to the tree-level $Z$-mediated cross-section. Therefore, this can
provide a rather stringent constraint on scenarios which, for example, avoid a coupling to the $Z$ by choice of the SU(2)$_L$ representation.
\item {\bf Anapole moment --} For charge-less Majorana fermions, there is a single nonzero multipole moment -  the anapole moment \cite{kayser} -- and in the 
scenarios considered here it can be quite sizable, consistent with the above result for the charge radius. As an explicit example, one may again have in 
mind a near-degenerate neutralino  and stau  in the MSSM. However, as the anapole is an axial moment, in elastic scattering it couples to the 
spatial nucleon current and so the cross-section is suppressed by ${\cal O}(v^2)\sim 10^{-6}$ \cite{ptv} which pushes it somewhat below the current
level of sensitivity.
\item {\bf (Chromo)electric polarizability --} The velocity suppression which afflicts the cross-section derived from the anapole moment does not apply to 
the polarizabilities. However, these operators require a chirality flip and so for resonant enhancement demand a double-degeneracy of the left- and
right-handed $\ch_2$ states with $\ch_1$. For weakly interacting states, e.g. $\tilde{\ta}_L$ and $\tilde{\ta}_R$ approximately degenerate with a neutralino (bino),
the diagram on the right of  Fig.~\ref{f5} leads to $\ch_E \sim 10^{-2} \al^2/( m_{\ch} (\De m)^2)$ in accord with the scaling of (\ref{scaling}). However, even in this
degenerate regime which resonantly enhances the amplitude, the cross-section for nucleon scattering is negligible due to the small electric energy
density in the nucleus $\langle N | E^2| N \rangle \sim 20$~MeV. However, there is an analogous chromoelectric polarizability, $\et_G$, replacing the 
photons in Fig.~\ref{f5}
with gluons which is far more significant since $\langle N | \al_s G^2 |N\rangle \sim m_N$. The corresponding contribution to the scattering cross-section 
in the resonant regime can be as large as $\si_{(p,n)}(\et_G) \sim 10^{-38}~{\rm cm}^2$ for $\De m \sim {\cal O}(100~{\rm MeV})$. This result is contained as
an appropriate limit of the general 1-loop results of Drees and Nojiri \cite{dn} for the MSSM, but is of less interest there as it requires two squarks nearly degenerate
with a neutralino LSP -- a somewhat unusual spectrum. 
\end{itemize}

For completeness, it is also worth remarking that if $\ch_1$ and $\ch_2$ are both fermionic as in models of type A, the existence of a light boson would allow the
amplitude to be independent of $m_\ch$. This follows from the same non-relativistic viewpoint as above, with the exception that now the heavy internal
line is fermionic and so the factors of $m_\ch$ cancel out. In practice, this situation is of less interest as there are no light scalars available to
provide the resonant diagram. However, this viewpoint does explain the corresponding results of \cite{hmns,strumia}, where the $W$-loop corrections
to the scattering cross-section were obtained and are explicitly independent of $m_{\rm DM}$ in the limit of large dark matter mass. 
This scaling is entirely consistent with the argument above, where the role of ``light boson'' is played by the $W$.
We note in passing that in the model of \cite{strumia} one should expect a significant difference in the scattering of Majorana 
and Dirac type dark matter. In the latter case, the cross section is enhanced due to a
significant contribution from the radiatively induced charge radius.  (See also a related discussion in Ref.~\cite{Essig}.)

 \section{Resonant WIMP annihilation}
 
 The pseudo-degeneracy in the WIMP sector was motivated in part by the 
generic need for enhanced annihilation mechanisms for heavy WIMPs. 
Although we have not discussed this in detail here, the present scenario  
with nearly degenerate charged states clearly leads to an interesting
coannihilation scenario for which the vector-mediated cross-section is 
further enhanced by the required resummation of Coulomb effects, as discussed recently
in \cite{coulomb}. Moreover, for small $\De m$ this will tend to a resonant process 
mediated by the formation of a metastable $(\ch_2^+\ch_2^-)$ state. 

\begin{figure}
\centerline{\includegraphics[bb=0 550 450 740, clip=true, width=11cm]{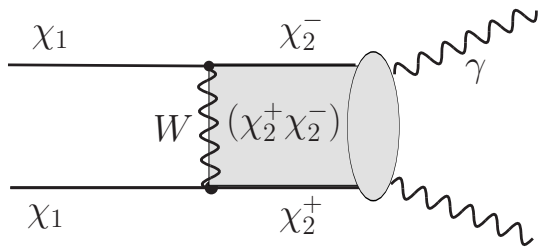}}
 \caption{\footnotesize A resonant contribution to the annihilation cross-section of $\ch_1\chi_1 \rightarrow 2\gamma$, mediated by the $(\ch_2^+\ch_2^-)$ bound state. }
\label{f6} 
\end{figure}

We will develop this possibility further in this section considering the following process
\be
\ch_1 + \ch_1 \to (\ch_2^+\ch_2^-) \to \gamma +\gamma,
\ee
which leads to the primary emission of monochromatic gamma rays inside {\em e.g.} the galactic center 
and can be effectively probed with existing and planned experiments. The Coulomb enhancements for
this process in the near-threshold regime were studied previously in \cite{hmns2}, but here we will focus
on the resonant contribution.

The binding energy of the $(\ch_2^+\ch_2^-)$ bound state is $\alpha^2m_\chi/4$. For TeV-scale WIMPs this is 
about 13 MeV, and thus can be quite comparable with the energy excess of $2\Delta m = 2(m_{\ch_2}- m_{\ch_1})$. 
In what follows, we analyze the regime where the $1S$ state for the $(\ch_2^+\ch_2^-)$ system is just above the 
energy threshold, and accessible via the kinetic energy of two WIMPs,
\be
0\leq 2\Delta m - \fr{m_\ch\alpha^2}{4} \la m_\ch v_m^2,
\ee
where $v_m$ is the maximal allowed WIMP velocity within the halo, on the order of $3\times 10^{-3}$. 
This kinematic regime is a narrow strip on the $(m_\ch,\Delta m)$ plane,
\be
0\leq 1.5 \times \left( \fr{\Delta m}{\rm 10~ MeV}\right) \times \left(\fr{\rm 1 ~TeV }{m_\ch}\right) -1 \la 0.7.
\ee

The spread in WIMP kinetic energy is on the order of an MeV and thus is much wider than 
both the capture width $\Gamma_{in}$ and the outgoing width $\Gamma_{out} = \Gamma_{2\gamma}$, 
so that the Breit-Wigner cross section is effectively a delta-function,
\be
\sigma = \fr{2\pi^2}{q_{cm}^2} g_* \Gamma_{2 \gamma} \delta(E-E_R),
\label{delta-f}
\ee
where $E_R= 2\Delta m - \fr{m_\ch\alpha^2}{4}$. 
In Eq.~(\ref{delta-f}) we took into account that the entrance 
width is much larger than $\Gamma_{2 \gamma}$, assuming no significant suppression of the WIMP-matter coupling,
and therefore the rate is controlled by the smallest width. The annihilation rate into 2$\gamma$'s of a bound state of two scalar 
particles is given by
\be
\Gamma_{2\gamma} = \fr{\alpha^5 m_\chi}{4} \simeq 5\times 10^{-6} \, {\rm MeV} \times \left(\fr{m_\ch}{\rm 1 ~TeV }\right) .
\label{2gamma}
\ee
For the parapositronium-like annihilation of two fermionic $\ch_2$'s, the rate is twice larger, and therefore the 
width $\Gamma_{2 \gamma}$ has rather minimal model-dependence. 

The capture rate is also straightforwardly calculable, but in contrast is significantly model-dependent. 
For models of Type A, the scaling of the entrance width is 
\be
\Gamma_{in}  \sim G_F^2 (\alpha m_\chi )^4 m_\chi \sqrt{\fr{8\Delta m}{m \alpha^2} -1} .
\ee
This rate is heavily dependent on $m_\chi$, but for most of the parameter  space relevant for WIMPs is larger than (\ref{2gamma}). 
The same holds true for capture in models of Type B. It is interesting to note that for capture processes
mediated by the Standard Model fermions, the $P$-wave capture may become significant. It is typically assumed that in the 
Galactic environment, where the velocity of colliding WIMPs is on the order of $10^{-3}$, $p$-wave annihilation is suppressed 
relative to the $s$-wave by a $v^2$ factor, or $10^{-6}$. In fact, the suppression factor is far less dramatic 
and for models with $\tau$-exchange it is $m_\ch^2v^2/m_\tau^2$, which is on the order of $10^{-2}$ for a 100 GeV WIMP. 
This allows for a sizable recombination rate of a neutralino pair with quantum numbers $L=1;~S=1;~J=0$
into a bound state of two staus. 

We are now ready to estimate the total annihilation rate by averaging (\ref{delta-f}) over the dark matter 
velocity distribution. If this distribution is approximated as Maxwellian,
then the rate is simply given by
\be
\langle \sigma v\rangle = \left( \fr{4\pi}{m_\ch T_{\rm eff}} \right)^{3/2} g_* \Gamma_{2 \gamma} \exp(-E_R/T_{\rm eff}),
\ee
where $T_{\rm eff}$ must be identified with the effective `temperature' of the WIMP gas, $\fr{3}{2}T_{\rm eff} 
\equiv \langle E \rangle$. For the typical velocity of dark matter inside the halo, this effective temperature is approximately
\be
T_{\rm eff} \simeq 300~{\rm keV} \times \left(\frac{m_\chi}{{\rm 1\,TeV}}\right).
\ee
 Inserting some characteristic numbers for a TeV mass WIMP, we obtain
\be
 \langle \sigma v\rangle \sim 5\times 10^{-34}\, {\rm cm}^2\times \left(\fr{\rm TeV}{m_\ch} \right)^2\exp\left(-E_R/T_{\rm eff} \right) \,,
 \label{annihilation_g}
\ee
which for $\De m$ tuned so that $E_R$ and $T_{\rm eff}$ are comparable, can be very large and indeed several orders of magnitude
larger than the characteristic mono-energetic $\gamma$ signal \cite{2gamma} in generic MSSM scenarios for example; similar enhancements
were obtained in \cite{hmns2} via a resummation of Coulomb effects near the bound-state threshold. 
Such mono-energetic
signals are far less sensitive to astrophysical backgrounds and so this regime is clearly promising for indirect searches.
Note that such a large galactic annihilation rate (\ref{annihilation_g}) is not in contradiction
with the required picobarn rate at freeze out, as this resonant effect is less important in the latter regime  due to
the larger effective temperature, which leads to an $O(10^{-2})-O(10^{-3})$ suppression in the 
thermal average.  Finally, we should note that this form of enhancement can also affect the production of charged states 
in the annihilation process, although the rate would be more model-dependent than for monochromatic photons.

\section{Concluding Remarks}

In this paper we have embarked on a study of some of the novel direct and indirect 
detection signatures of pseudo-degenerate WIMP dark matter.
Although this scenario lies within the characteristic WIMP sector of viable supersymmetric models 
and is thus far from exotic,  the MeV-scale degeneracy
in the spectrum does lead to some rather exotic signatures; most prominently through recombination processes with nuclei. 
In this concluding section,
we will briefly summarize the results and comment on some further probes of this scenario.

Although we have refrained from performing detailed calculations of the capture rates, 
and our treatment of the nuclear physics was somewhat cursory, the estimates
obtained in Section~3 are sufficient to draw some broad conclusions on direct detection to focus future studies. 
For models of Type A with a weak current, 
we found that searches for anomalous heavy isotopes in fact do not impose any significant
 constraints on the inelastic cross-section with nuclei. However, our estimate for
the rate was relatively small, at best comparable with the existing direct detection limit on the elastic cross-section. 
For models of Type B, the cross-section
is constrained by heavy isotope searches, but only in a rather small range arising from binding with light elements. 
For splittings in the range (\ref{recom_range1}),
the constraints are far weaker and our estimate for the capture rate suggests that simple models could 
produce rates a few orders of magnitude larger than
the existing constraint on the elastic cross-section. Clearly it would be very interesting to 
explore this possibility further and to determine whether conventional
direct detection experiments could probe capture processes of this type.

A significant motivating factor in the analysis of this scenario was that the novel direct detection signatures may imply somewhat weaker constraints on
the cross-section with matter than would apply to generic WIMPs. This opens up the possibility of other interesting indirect detection possibilities in galactic astrophysics.
A possible signal of this type, the observed excess of 511 keV photons from the galactic center, was considered recently within this framework \cite{pr1} but
seemingly required too large a cross-section to be consistent with terrestrial heavy isotope bounds (in this case the mass splitting required was ${\cal O}$(1 MeV)
which allows for binding with C, N etc). Here we have explored several other signatures for indirect detection.
We  noted that the possibility of observing highly energetic neutrinos from DM annihilation in the sun is significantly diminished within these scenarios, due to 
the far higher rate for the WIMPs to be captured by iron than to annihilate. If the mass
splitting is tuned appropriately, we also observed that the annihilation rate to two mono-energetic $\gamma$'s, e.g. in the galactic center, could be resonantly enhanced by
several orders of magnitude through an intermediate $(\ch_2^+\chi_2^-)$ bound state. 

It would clearly be interesting to explore other indirect signatures beyond the traditional $\gamma$-signal from WIMP annihilation, as may arise through
the possibility for recombination processes in astrophysical environments which may lead to new detection strategies, or alternatively more stringent
constraints to be placed on these scenarios. Finally, we should emphasize that we have focused here on direct and indirect detection of WIMPs in the galactic halo, 
and have avoided the question of direct production in colliders. This is primarily because direct production would at least in the near term be limited to the low 
mass regime, which contrasts somewhat with part of the motivation for the pseudo-degenerate scenario, which was the need for heavier states. Nonetheless, this 
question is certainly of general interest and existing analyses of the collider physics of the WMAP strip region in the CMSSM should include the regime
where the neutralino LSP is a pseudo-degenerate WIMP. In this context, the charged $\ch_2^{\pm}$ states could be sufficiently long lived to either escape or become
bound inside the detector.

\subsection*{Acknowledgments}

We are grateful to Mark Boulay for a helpful discussion on WIMP direct detection. This work was supported in part by NSERC, Canada, and research at the Perimeter Institute
is supported in part by the government of Canada through NSERC and by the province of Ontario through MEDT.

\end{document}